\documentclass[twocolumn,showpacs,preprintnumbers,amsmath,amssymb,prl]{revtex4}

\usepackage{graphicx}
\usepackage{dcolumn}
\usepackage{bm}
\usepackage{color}
\usepackage{soul}

\setstcolor{red}

\newcommand{\rmi}{{\rm i}}

\begin{document}

\title{Electric-field-driven resistive switching in dissipative Hubbard model}

\author{Jiajun Li$^1$, Camille Aron$^{2,3}$, Gabriel
Kotliar$^2$ and Jong E. Han$^1$}
\affiliation{
$^1$ Department of Physics, State University of New York at Buffalo, Buffalo,
New York 14260, USA\\
$^2$ Department of Physics, Rutgers University, New Jersey 08854, USA\\
$^3$ Department of Electrical Engineering, Princeton University, New
Jersey 08455, USA.
}

\date{\today}

\begin{abstract} 

We study how strongly correlated electrons on a dissipative lattice
evolve from equilibrium under a constant electric field,
focusing on the extent of the linear regime and hysteretic non-linear
effects at higher fields. We access the non-equilibrium steady states,
non-perturbatively in both the field and the electronic interactions, by
means of a non-equilibrium dynamical mean-field theory in the Coulomb
gauge. The linear response regime, limited by Joule heating,
breaks down at fields much smaller than the
quasi-particle energy scale. For large electronic interactions, strong
but experimentally accessible electric fields can induce a resistive
switching by driving the strongly correlated metal into a Mott
insulator. We predict a non-monotonic upper switching field due to an interplay of particle renormalization and the field-driven temperature.
Hysteretic $I$-$V$ curves suggest that the non-equilibrium
current is carried through a spatially inhomogeneous metal-insulator
mixed state.

\end{abstract}


\maketitle

Understanding of solids driven  out of equilibrium by  external
fields~\cite{kadanoff,mahan} has been one of the central goals
in condensed matter physics for the past century and is very relevant
to nanotechnology applications such as resistive transitions.
Multiple studies of this phenomenon have been performed in 
semiconductors and oxides~\cite{French,Tokura,Parkin,Doug,noh,VO2-Joule-Adam,VO2-Joule-Basov,
VO2-Joule-Schuller}.
In oxides, the application of an
electric field can lead to a dramatic drop of resistivity up to 5 orders
of magnitude. The relatively accessible threshold fields $E_{\rm th}
\sim 10^{4-6}$~V/m and the hysteretic $I$-$V$ curves make them good candidates for
the fabrication of novel electronic memories.
A Landau-Zener type of mechanism~\cite{Oka} seems unlikely as it
predicts a threshold field  on the order of $10^{8-9}$ V/m. In narrow
gap chalcogenide Mott insulators, 
an avalanche breakdown was suggested with
$E_{\rm th} \sim E_{\rm gap}^{2.5}$~\cite{French}.
Yet, the resistive switchings in other classes of correlated materials
do not seem to involve solely electronic mechanisms.
In organic charge-transfer complexes, it is believed to occur
\textit{via} the electro-chemical migration of
ions~\cite{Tokura,Parkin}.
Finally, there are strong indications that a Joule heating
mechanism occurs in some binary oxides such as NiO~\cite{noh} and
VO$_2$~\cite{VO2-Joule-Adam,VO2-Joule-Basov,VO2-Joule-Schuller}: the
electric-field-driven current locally heats up the sample which
experiences a temperature-driven resistive switching.

These experiments raise basic questions of how a strongly correlated
state \textit{continuously} evolves out of equilibrium under an external
field, and how we describe the non-equilibrium steady states that
consequently emerge. We develope a much needed basic \emph{microscopic}
theory of the driven metal-insulator transition.

Building on earlier theoretical efforts~\cite{turkowski,
freericks,joura,eckstein,Oka,aoki,vidmar,aron,amaricci,aron1,mitra,diss1,diss2,neqdmft,nagaosa,okamoto,fabrizio}
we identify in  a canonical model of strongly interacting electrons a
region where electric-field-driven resistive switching takes place.  We
demonstrate how Joule heating effects modify the linear response regime
and how, away from the linear regime, the same Joule physics leads to
the hysteretic resistive transitions of the strongly correlated system.
The derived energy scales for resistive transitions are orders of
magnitude smaller than bare model parameters, within the feasible
experimental range.

We study the  Hubbard model in a constant and homogeneous electric field
$\mathbf{E}$ which induces electric current $\mathbf{J}$. After a
transient regime, a non-equilibrium steady state establishes if the
power injected in the system, $\mathbf{J}\cdot \mathbf{E}$, is balanced
by coupling the system to a thermostat which can absorb the excess of
energy \textit{via} heat
transfer~\cite{aron,amaricci,aoki,vidmar,diss1,diss2}.  
The thermostat is modeled by
identical fermion reservoirs attached to each tight-binding (TB) sites.  In the Coulomb
gauge, the electric field amounts in an electrostatic potential $-\ell
E$ imposed on the $\ell$-th TB site ($\ell=-\infty,\cdots,\infty$) and
on its associated fermion bath~\cite{diss2}. The model is fully
consistent with gauge-covariant models~\cite{aron}. The non-interacting
Hamiltonian reads, \begin{align} \hat{H}_0 & =
-\gamma\sum_{\ell\sigma}(d^\dagger_{\ell+1,\sigma}d_{\ell\sigma}+{\rm
H.c.})
-\frac{g}{\sqrt{V}}\sum_{\ell\alpha\sigma}(d^\dagger_{\ell\sigma}c_{\ell\alpha\sigma}
+{\rm H.c})\nonumber \\ &+\sum_{\ell\alpha\sigma}\epsilon_\alpha
c^\dagger_{\ell\alpha\sigma}c_{\ell\alpha\sigma}
\!-\!\sum_{\ell\sigma}\ell E(d^\dagger_{\ell\sigma}d_{\ell\sigma}
\!+\!\sum_\alpha \! c^\dagger_{\ell\alpha\sigma}c_{\ell\alpha\sigma}),
\label{eq:h0} \end{align} where $d^\dagger_{\ell\sigma}$ are the
tight-binding electron creation operators at the $\ell$-th site with
spin $\sigma = \uparrow$ or $\downarrow$, and
$c^\dagger_{\ell\alpha\sigma}$ are the corresponding reservoir electron
operators attached.  $\alpha$ is a continuum index corresponding to the
reservoir dispersion relation $\epsilon_\alpha$ defined with respect to
the electrostatic potential $-\ell E$.  $g$ is the overlap between the
TB chain and the reservoirs of length $V$ which will be sent to
infinity, assuming furthermore that the reservoirs remain in equilibrium
at bath temperature $T_{\rm b}$.  Later we will extend this chain  into
higer dimensional lattice.  The electric field does not act within each
reservoirs whose role is to extract energy but not electric charge from
the system~\cite{diss2}.  We use a flat density of states (infinite
bandwidth) for the reservoir spectra $\epsilon_\alpha$, and define the
damping parameter as $\Gamma=V^{-1}\pi
g^2\sum_\alpha\delta(\epsilon_\alpha)$.  We work with $\hbar=e=k_{\rm
B}=a=1$ in which $e$ is the electronic charge and $a$ is the lattice
constant.  In the rest of this Letter, we measure energies in units of
the full TB bandwidth $W=4\gamma=1$ (1-$d$) and $W=12\gamma=1$ (3-$d$).
The exact solution of the non-interacting model in Eq.~(\ref{eq:h0}) has
been shown~\cite{diss1,diss2} to reproduce the conventional Boltzmann
transport theory despite the lack of momentum transfer scattering.  The
Hubbard model $\hat{H}=\hat{H}_0+\hat{H}_1$ is defined with the on-site
Coulomb interaction parameter $U$ as
\begin{equation}
\hat{H}_1=U\sum_\ell\left(d^\dagger_{\ell\uparrow}d_{\ell\uparrow}-\frac12\right)
\left(d^\dagger_{\ell\downarrow}d_{\ell\downarrow}-\frac12\right).
\end{equation}
Our calculations are in the particle-hole symmetric limit.

We use the dynamical mean-field theory (DMFT~\cite{dmft,neqdmft}) to treat the many-body
interaction \textit{via} a self-consistent local approximation of the self-energies.
Note that the self-energy has contributions from both the many-body interaction $\hat{H}_1$ and the coupling to the reservoirs: $\Sigma^{r}_{\rm tot}(\omega)=-\rmi \Gamma+\Sigma^r_U(\omega)$ and $\Sigma^<_{\rm tot}(\omega)=2\rmi\Gamma f_{\rm FD}(\omega)+\Sigma^<_U(\omega)$ with the Fermi-Dirac (FD) distribution $f_{\rm
FD}(\omega)\equiv[1+\exp(\omega/T_{\rm b})]^{-1}$. Once the local retarded and lesser self-energies are computed, one can access the full retarded and lesser Green's functions (GFs).
Note that in a homogeneous non-equilibrium steady state, all the TB
sites are equivalent. In the Coulomb gauge, this leads to
$G^{r,<}_{\ell\ell'}(\omega)=G^{r,<}_{\ell+k,\ell'+k}(\omega+kE)$ and
similarly for the self-energies~\cite{diss2,okamoto}, 
as can be derived via a gauge transformation from the temporal gauge.

Below, we present the implementation of our DMFT scheme in
the Coulomb gauge directly in the steady states. It
consists in singling out one TB site -- say $\ell=0$ -- (often referred
as impurity) and replacing its direct environment (\textit{i.e.} 
semi-infinite dissipative Hubbard chains and its own reservoir) with a self-consistently determined non-interacting environment (often referred as Weiss ``fields''). The local electronic problem is then treated by means of an impurity solver.

For given self-energy
[$\Sigma^{r,<}_{\ell}(\omega)\equiv\Sigma^{r,<}_U(\omega+\ell E)$], the on-site Green's functions
obey the following Dyson equations
 \begin{eqnarray}
G^r(\omega)^{-1} & = & \omega-\Sigma^r_{\rm tot}(\omega)
-\gamma^2F^r_{\rm tot}(\omega),\\
G^<(\omega) & = & |G^r(\omega)|^2[\Sigma^<_{\rm tot}(\omega)
+\gamma^2F^<_{\rm tot}(\omega)],
\end{eqnarray}
in which $\gamma^2 F^{r,<}_{\rm tot}$ are the total hybridization functions to the left and right semi-infinite chains, $F^{r,<}_{\rm tot}(\omega) =F^{r,<}_+(\omega+E) +F^{r,<}_-(\omega-E)$.
$F_+(\omega)$ is the on-site retarded GF at the end of the RHS-chain ($\ell=1$) which obeys the self-similar Dyson equation 
\begin{equation}
F^r_+(\omega)^{-1} =
\omega-\Sigma^r_{\rm tot}(\omega)-\gamma^2F^r_+(\omega+E),
\end{equation}
which can be solved recursively after more than 500 iterations. $F_-(\omega)$ corresponds to the GF of the LHS-chain and can be obtained similarly. The non-interacting parts of the impurity GFs, ${\cal G}$, are constructed using
\begin{eqnarray}
{\cal G}^r(\omega)^{-1} & = & \omega+\rmi\Gamma
-\gamma^2F^r_{\rm tot}(\omega)\\
{\cal G}^<(\omega) & = & |{\cal G}^r(\omega)|^2[2\rmi\Gamma f_{\rm FD}(\omega)
+\gamma^2F^<_{\rm tot}(\omega)].
\end{eqnarray}
The local self-energies are obtained by means of the iterative-perturbation theory
(IPT) up to the second-order in the Coulomb parameter $U$:
$\Sigma^{\gtrless}_U(t)=U^2[{\cal G}^\gtrless(t)]^2{\cal
G}^\lessgtr(t)$. The GFs are updated with this self-energy using the above Dyson's equations and the procedure is repeated until convergence is achieved.

We generalize the above method to higher dimensions.
With the electric-field along
the principal axis direction, ${\bf E}=E\hat{\bf x}$,
the lattice is translation invariant in the
perpendicular direction and the above construction of the Dyson's
equation can be carried out independently per each perpendicular
momentum vector. See Supplementary Material for a detailed discussion.
Below, we present results of the model in one and three dimensions.

We first discuss the linear response regime. 
Within the DMFT, the DC conductivity in the limit of zero temperature
and zero electric field can be obtained via the Kubo formula as $
\sigma_{\rm DC} \propto \lim_{\omega\to 0}\sum_{\bf k} \int d\nu
\rho_{\bf k}(\nu)
\rho_{\bf k}(\nu+\omega)[f_{\rm FD}(\nu)-f_{\rm FD}(\nu+\omega)]/\omega
=\sum_{\bf k} \int d\nu [\rho_{\bf k}(\nu)]^2\delta(\nu)$ with the spectral function
at a given wave-vector ${\bf k}$ $\rho_{\bf k}(\nu)=-\pi^{-1}{\rm
Im}[\nu-\epsilon_{\bf k}+\rmi\Gamma-\Sigma^r_U(\nu)]^{-1}$. 
Therefore, 
as long as $\Sigma^r_U(\nu)\to 0$ as $\nu\to
0,T\to 0$, the DC conductivity is independent of the interaction.
This argument is similar to the one used by Prange and Kadanoff~\cite{prange}
for  the electron-phonon interaction. 
Recent calculations did not have  access to the linear response 
regime~\cite{aoki,aron,amaricci}.

\begin{figure}
\rotatebox{0}{\resizebox{1.6in}{!}{\includegraphics{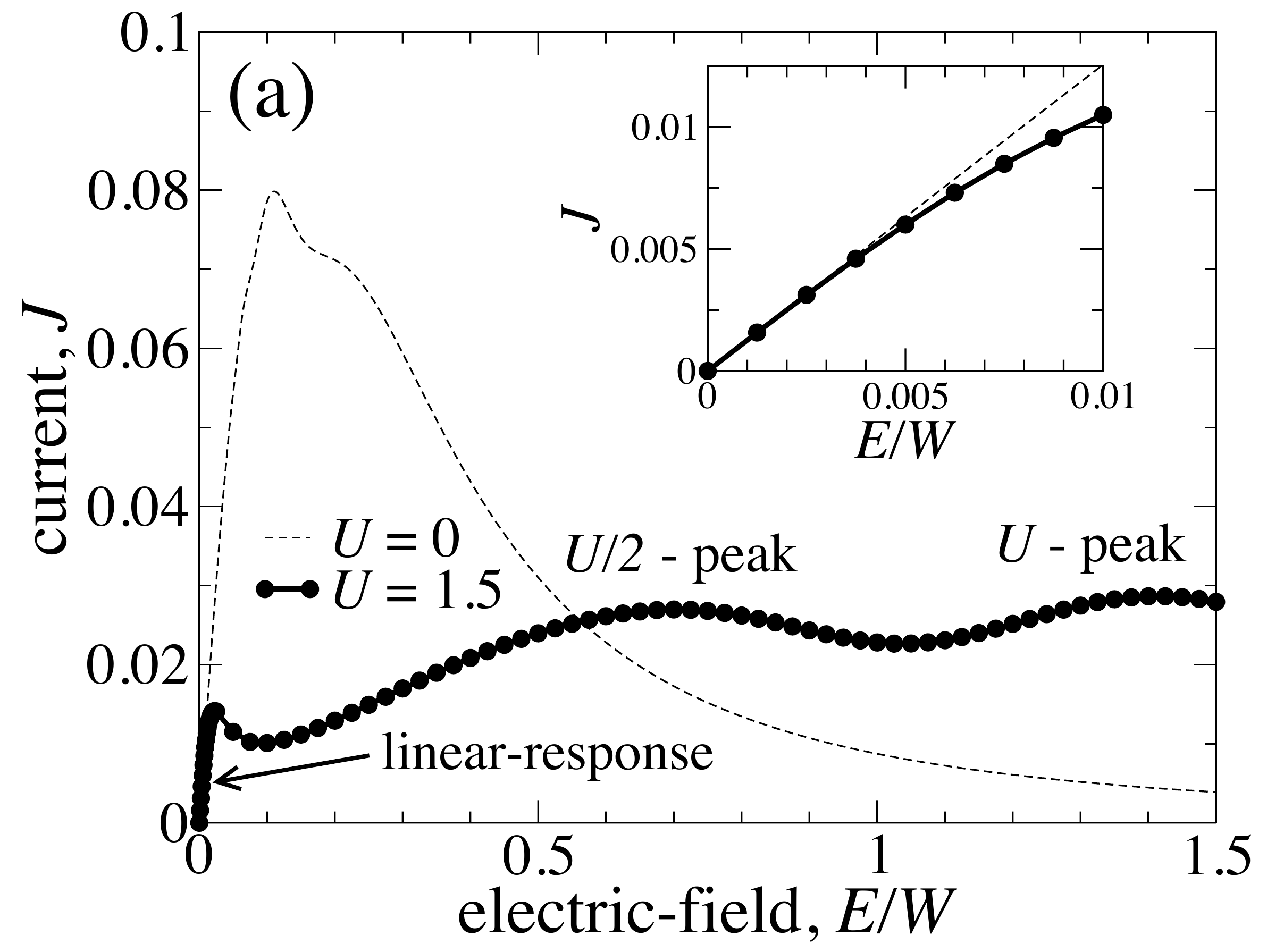}}}
\rotatebox{0}{\resizebox{1.6in}{!}{\includegraphics{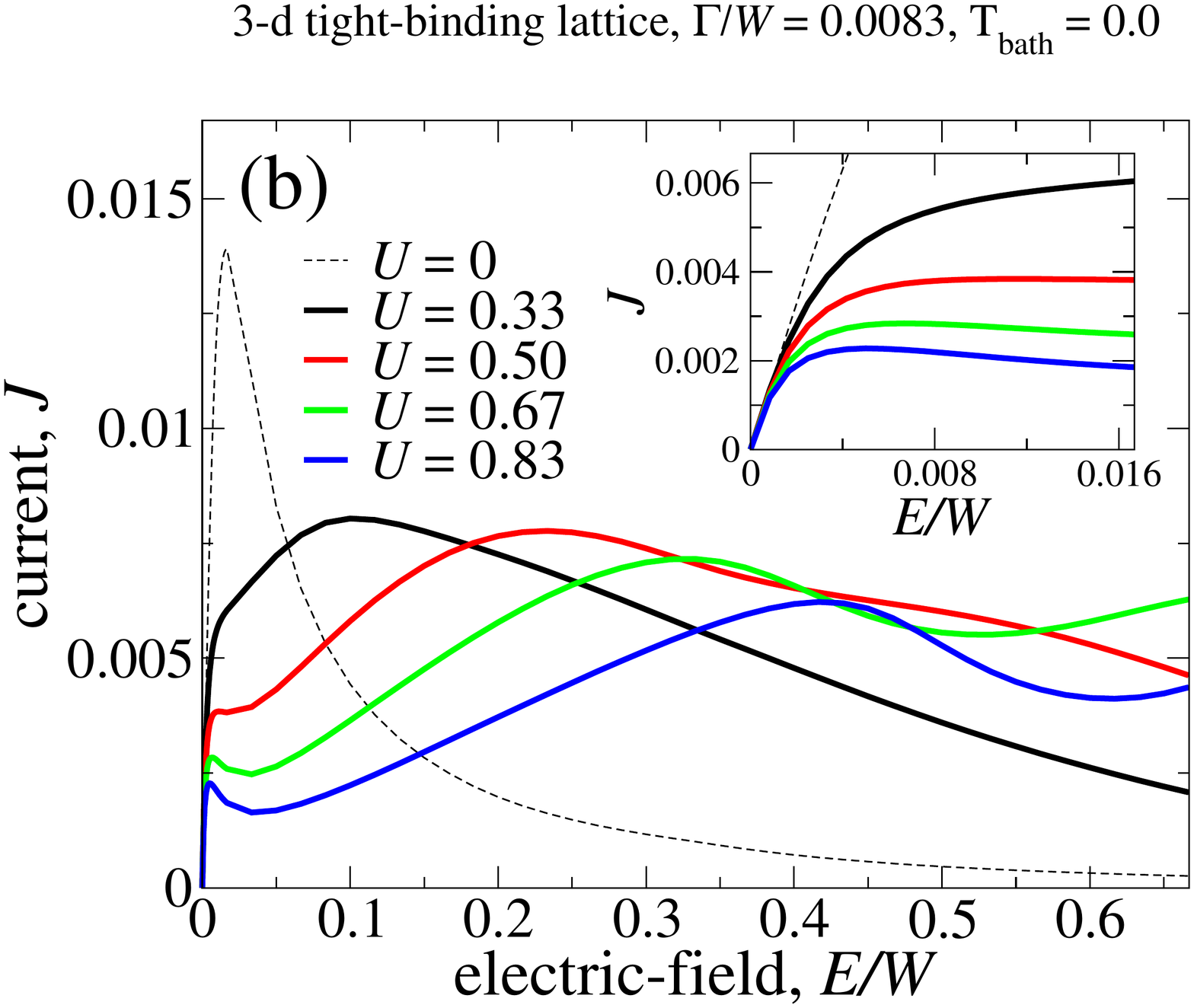}}}
\caption{(color online) Electric current (per spin)
$J$ vs. electric field $E$. (a)
1-$d$ chain with damping $\Gamma=0.0625W$ and fermion bath temperature
$T_{\rm b}=0.00125W$ with the 1-$d$ TB bandwidth $W=4\gamma$. 
The linear conductance in the small field limit (magnified in the inset) is the same for non-interacting
($U=0$) and interacting ($U=1.5W$) models. After the 
conductivity deviates from the linear response behavior, inelastic
contributions appear at $E=U/2$ and $E=U$.
(b) 3-$d$ lattice with $\Gamma=0.0083W$ and $T_{\rm b}=0.00042W$ with
the 3-$d$ TB bandwidth $W=12\gamma$. The main features remain similar to the 1-$d$ case.
All following energies are in unit of $W$, unless otherwise mentioned.
}
\label{fig1}
\end{figure}

FIG.~\ref{fig1} confirms the validity of the linear response analysis.
The initial slope of the $J-E$ relation is
independent of the interaction strength $U$~\cite{fabrizio} both in (a) one and (b)
three-dimension. The linear behavior deviates at the field $E_{\rm
lin}\approx 0.003$ in (a), orders of magnitude smaller than the
renormalized bandwidth
$W^*=zW\approx 0.5$ with the
equilibrium renormalization factor $z=[1-{\rm
Re}\partial\Sigma^r_U(\omega)/\partial\omega]^{-1}_{\omega=E=T_{\rm b}=0}$. 

With increasing E-field, the contribution at $E=U/2$ is a two-step
resonant process which can be viewed as a consequence of the energy
overlap between the lower/upper Hubbard bands of the left/right
neighboring sites with the in-gap states present at the Fermi
level~\cite{aron1}. The current peak at $E=U$ is due to the direct
overlap of the Hubbard bands on neighboring sites~\cite{joura,aron1}.

\begin{figure}
\rotatebox{0}{\resizebox{1.7in}{!}{\includegraphics{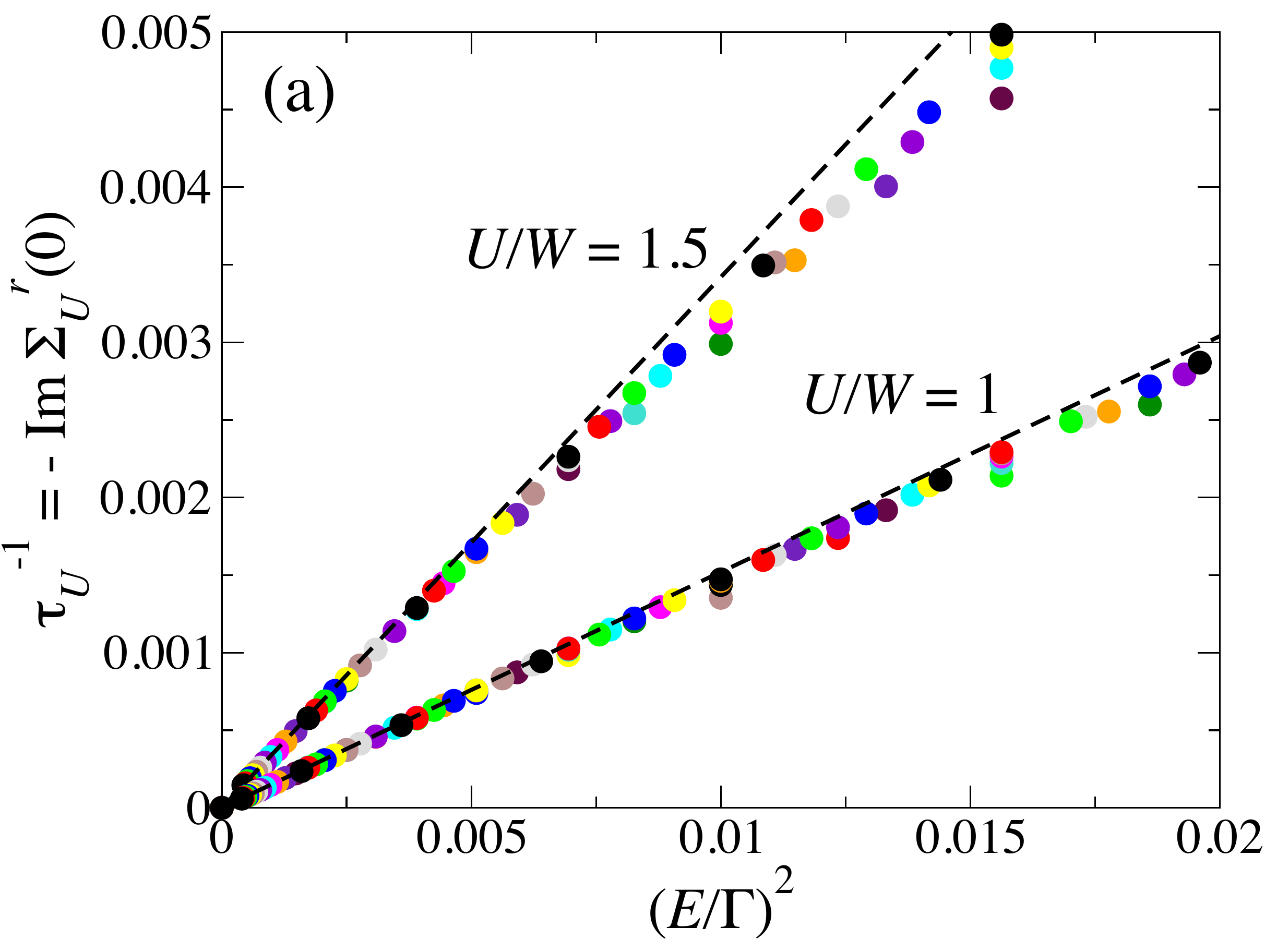}}}
\rotatebox{0}{\resizebox{1.6in}{!}{\includegraphics{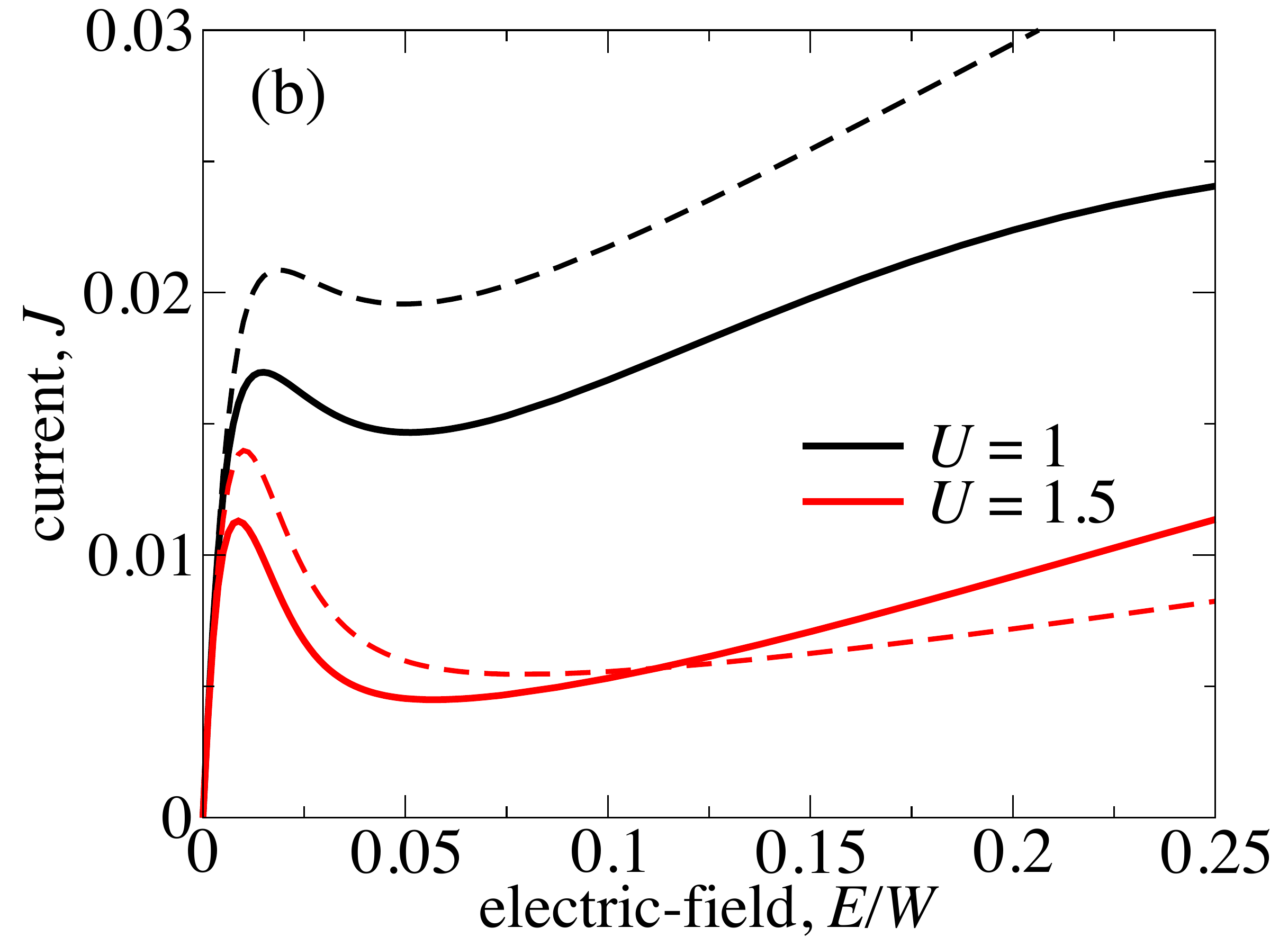}}}
\caption{(color online) (a) Interacting scattering rate, $\tau^{-1}_U=-{\rm
Im}\Sigma^r_U(\omega=0)$, plotted against $(E/\Gamma)^2$.
Different colors denote different damping $\Gamma=0.0125,\cdots,0.06$ with
the interval of 0.0025. For small $(E/\Gamma)$, the numerical results on the 1-$d$ chain
collapse on well-defined lines at $U=1$ and $1.5$. The dashed lines
are predictions based on the \textit{equilibrium} self-energy with the
temperature replaced by the non-interacting effective temperature $T_{\rm eff}$ given in Eq.~(\ref{teff}).
The remarkable agreement proves that Joule heating controls the scattering in the small field
limit. (b) Comparison of the current and the Drude
formula estimate with the total scattering rate
$\Gamma+\tau_U^{-1}$, with qualitative agreement beyond the linear
response limit.
}
\label{fig2}
\end{figure}
 
The immediate departure from the linear conductivity at very small
fields can be well understood with a Joule heating scenario in which the Coulombic interaction is the dominant scattering process and is rapidly modified by an increasing effective temperature as the field is increased.
We first demonstrate this effective temperature effect by showing in Fig.~\ref{fig2}(a) that the scattering rates from the Coulomb interaction, $\tau^{-1}_U=-{\rm
Im}\Sigma^r_U(\omega=0)$, for different sets of the damping $\Gamma$
collapse onto a scaling curve as a function of $(E/\Gamma)^2$ for small $E$. This scaling is clearly evocative of the well known $T^2$ behavior of equilibrium retarded self-energies.

In the non-interacting 1-$d$ chain with $T_{\rm b}=0$, the effective temperature has been obtained in the small field limit as~\cite{diss2,mitra}
\begin{equation}
T_{\rm eff}=\frac{\sqrt{6}}{\pi} \, \gamma \, \frac{E}{\Gamma}\,.
\label{teff}
\end{equation}
Inserting this $T_{\rm eff}$ into the \textit{equilibrium} perturbative
self-energy~\cite{yamada}, we obtain in the weak-$U$ limit
\begin{equation}
\tau_{U}^{-1} = -{\rm Im}\Sigma^r_{\rm eq}(\omega=0,T_{\rm eff})
\approx \frac{\pi^3}{2} \, A_0(0)^3 \, U^2 \, T_{\rm eff}^2,
\label{eq:scatt}
\end{equation}
which is represented by the dashed lines in Fig.~\ref{fig2}(a). Here
$A_0(0)=(\pi\sqrt{\Gamma^2+4\gamma^2})^{-1}$ is the non-interacting DOS
at $\omega=0$. The robust agreement in the self-energies leaves no doubt that the electron
scattering is dominated by the Joule heating
with $T_{\rm eff}$ given
with Eq.~(\ref{teff}) in the linear response limit in the presence of
interaction. $T_{\rm eff}$ then deviates strongly from this behavior
outside the narrow linear regime, as discussed below.

The scattering rate can be directly related to the electric current via
the Drude conductivity $J(E)=\sigma_{\rm DC}(E)E$
with the non-linear DC conductivity $\sigma_{\rm
DC}(E)$. In
the non-interacting limit, the linear conductivity can be written
as $\sigma_{0,\rm DC}=2\gamma^2/(\pi\Gamma\sqrt{\Gamma^2+4\gamma^2})$~\cite{diss2}.
In FIG.~\ref{fig2}(b), we plot the Drude formula with the scattering rate $\Gamma$ replaced by the
total scattering $\Gamma_{\rm tot}=\Gamma+\tau_U^{-1}$. The qualitative agreement with the numerical results extends over a wide range of the $E$-field, well beyond the linear regime. 

Using Eq.~(\ref{eq:scatt}), the current at small field can be
approximated as $J=\sigma_{0,DC}E/(1+E^2/E_{\rm lin}^2)$ with
the departure from the linear behavior occuring around
(from the condition $\Gamma\approx\tau_U^{-1}$ at
$E=E_{\rm lin}$), 
$
E_{\rm lin}\approx(8\pi^2/3)^{1/2}\gamma^{1/2}\Gamma^{3/2}/U.
$
This estimate is valid away from $U=0$ and the metal-insulator limit,
and agrees well with FIG.~\ref{fig2}(b)~\cite{impurity}.
We emphasize that, while negative-differential-resistance (NDR) behaviors occur
typically in periodic structures due to the Bloch
oscillations~\cite{lebwohl} as the
dashed lines ($U=0$) in Fig.~\ref{fig1}, the NDR here 
comes from strong non-linear scattering enhanced by the Joule heating. 

\begin{figure}
\rotatebox{0}{\resizebox{3.3in}{!}{\includegraphics{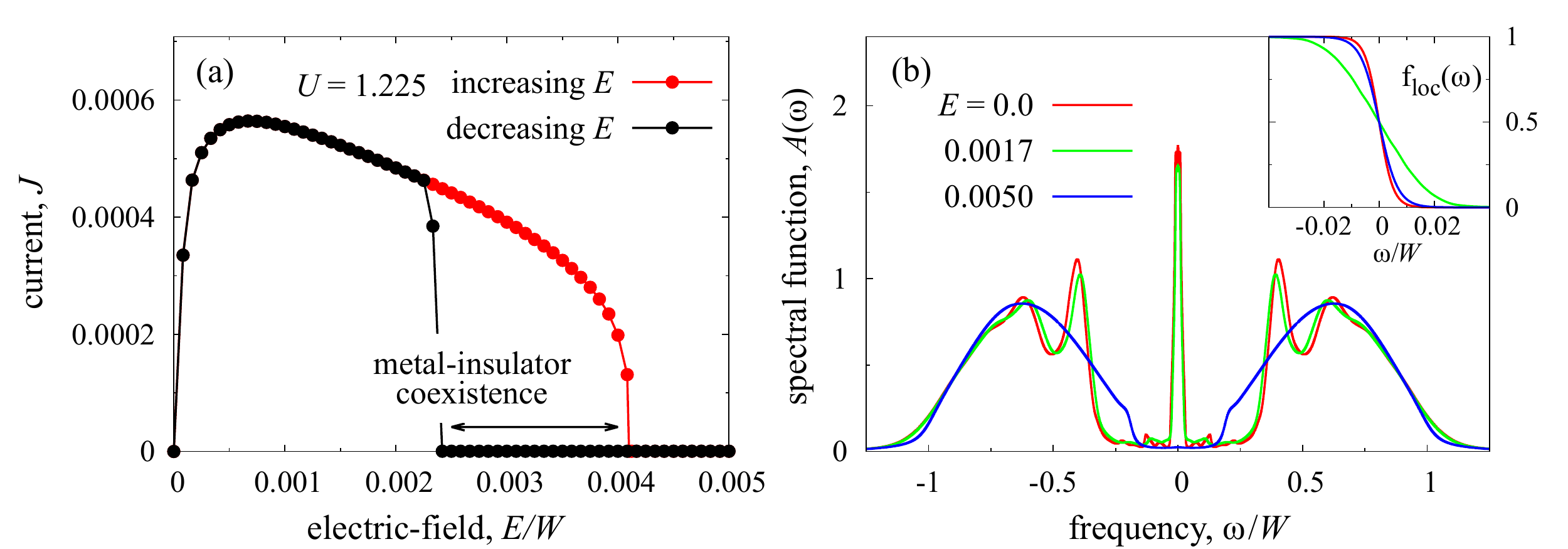}}}
\caption{(color online)
(a) Electric-field driven metal-to-insulator transition (MIT) in the
vicinity of a Mott-insulator at $U=1.225$, $\Gamma=0.00167$ and $T_{
b}=0.0025$ in a 3-dimensional cubic lattice with electric field
in $x$-direction. The metallic state at zero field becomes insulating
at electric field of magnitude orders of magnitude smaller than bare
energy scales. Depending on whether the electric-field is
increased or decreased, metal-insulator hysteresis occurs with a
window for phase-coexistence. (b) Spectral function and distribution
function $f_{\rm loc}(\omega)$ with increasing electric-field. The
quasi-particle (QP) spectral weight rapidly disappears near the MIT driven
by the electric-field, opening
an insulating gap. The non-equilibrium energy distribution function indicates that the system
undergoes a highly non-monotonic cold-hot-cold temperature evolution near the MIT.
}
\label{fig3}
\end{figure}

In the presence of weak dissipation and strong electronic interactions,
the non-equilibrium evolution becomes more dramatic. With the effective
temperature, Eq.~(\ref{teff}), having a singular limit as $\Gamma\to 0$,
the electron temperature tends to rise very sharply as the field is
applied. This effect, together with a small value of the renormalized coherent energy scales, causes the system to immediately deviate from the linear response regime,
preventing itself from
overheating. This mechanism, in a vicinity of a quantum phase
transition, can strongly modify the state of a system.
Indeed, we will show that there is a region of the parameters $U$ and $E$ for which the non-equilibrium Dyson's equations have two distinct solutions, one corresponding to an incoherent metal and the other to an insulator.

 In Fig.~\ref{fig3}(a), we
start from a metallic state at $U=1.225$, and increase the electric-field
from zero. We use the self-consistent solution
at a certain $E$-field as an input to the next $E$ run.
As discussed above, the system has an extremely narrow linear response window with
$E_{\rm lin} \sim 10^{-4}$, followed by an NDR behavior. 
As the electric-field is further increased, an
electric-field-driven metal-to-insulator RS occurs at $E_{\rm MIT}\approx
0.004$. Similar strong non-linear $I$-$V$ behavior
followed by a resistive transition has been observed in NiO~\cite{noh}.
After gradual changes in the spectral functions in Fig.~\ref{fig3}(b),
a finite insulating gap opens abruptly after the RS. The local energy distribution function
$f_{\rm loc}(\omega)$, defined as $f_{\rm loc}(\omega)=-\frac12 {\rm
Im}G^<(\omega)/{\rm Im}G^r(\omega)$, evolves from the FD function at
zero field to a shape with a high effective temperature.
At the RS, the Joule heating nearly stops and the TB lattice goes back to the low
temperature state~\cite{lowT}.
We emphasize that the energy scale hierarchy
\begin{equation}
E_{\rm lin}\ll E_{\rm MIT}\ll W^*
\end{equation}
observed above
differs markedly from that in the quantum dot transport~\cite{kondo} in
which the dissipation occurs outside the quantum dot region and the
bias scale for decoherence is comparable to the QP energy scale.

\begin{figure}
\rotatebox{0}{\resizebox{3.3in}{!}{\includegraphics{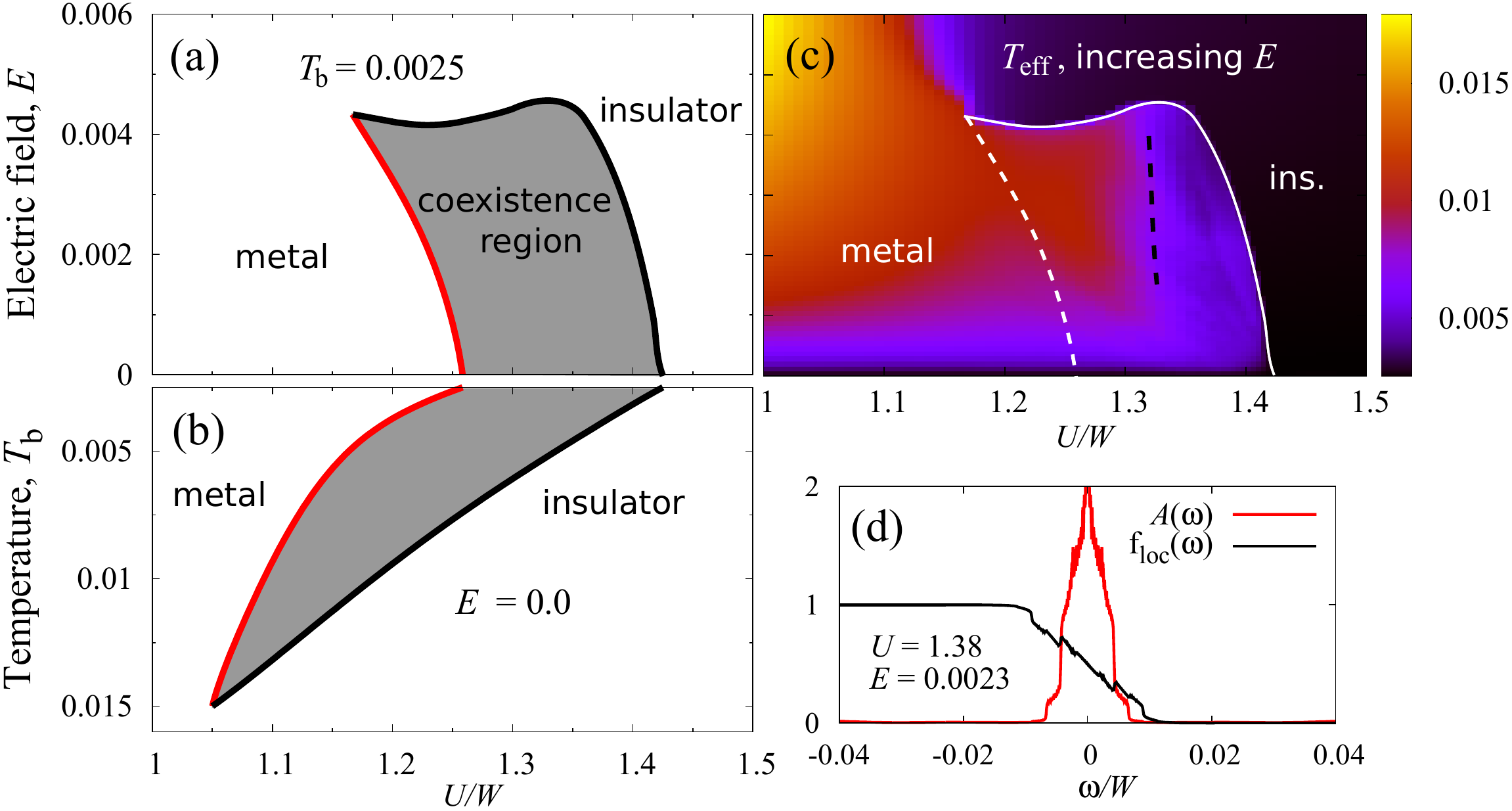}}}
\caption{(color online) Phase diagram of metal-insulator transition in
a cubic lattice driven by (a) electric field and (b) temperature. The
metal-insulator coexistent phase exists between the metal-to-insulator
transition
(black line) with increasing $E$ or $T_{\rm b}$, and the
insulator-to-metal transition
(red line) with decreasing $E$ or $T_{\rm b}$.
$\Gamma=0.00167$.
(c) Effective temperature $T_{\rm eff}$ map with increasing $E$,
with the white line for the MIT. The white dashed line becomes the phase
boundary with decreasing field.
(d) Spectral and distribution functions for strong $U$ beyond the
crossover line [black dashed in (c)]. Quasi-particle states are
disconnected from incoherent spectra and their statistical property becomes
strongly non-thermal.
}
\label{fig4}
\end{figure}

Fig.~\ref{fig4}(a-b) show the metal-insulator coexistence.  Our
estimate of the threshold electric field $E_{\rm MIT}\approx 0.004$ at
$U=1.225$ can be converted to $E_{\rm MIT}=10^7-10^8$ V/m if $U=1-10$ eV.  
Based on the balance between the Joule heating and the
dissipation~\cite{diss2,altshuler},
a scaling argument~\cite{supplement} implies that
the critical field decreases with damping as $E_{\rm
MIT}\propto\sqrt{\Gamma}$.  Therefore, accounting for the range of
experimental threshold fields would require $\Gamma$
on the order of $10^{-3}$~meV. 
We stress that the model successfully
captures, at a microscopic level, the qualitative features of the
resistive switching phenomenon but a more quantitative analysis calls
for a better modelling of the dissipative mechanisms.

While the phase diagram for the RS of Fig.~\ref{fig4}(a) generally reflects that of
the equilibrium MIT~\cite{dmft} in (b), the upturn of the upper
critical E-field (black line) in Fig.~\ref{fig4}(a) with increasing $U$
is counter-intuitive.
This originates from an interplay of different
scaling regimes for large and small $U$ separated by the crossover line
(dashed line) at about $U_{\rm cross}/W\approx 1.32$. For small $U<U_{\rm
cross}$, the QP bandwidth $W^*$ is larger than $T_{\rm eff}$ and the
scaling relation $T_{\rm eff}\propto\sqrt{E/U}$~\cite{supplement}
results well away from the linear regime, Eq.~(\ref{teff}). 
However, for $U>U_{\rm cross}$ with
$W^*\lesssim T_{\rm eff}$, $T_{\rm eff}$ increases with $E$ much
weakly~\cite{supplement}, as seen in Fig.~\ref{fig4}(c).
This slow increase of $T_{\rm eff}$ allows a larger
critical field and leads to the maximum $E_{\rm MIT}(U)$ near $U=U_{\rm
cross}$ -- a prediction which can be experimentally verified. The spectral and
distribution functions in Fig.~\ref{fig4}(d) for $U>U_{\rm cross}$, show
the QP states spectrally disconnected incoherent electrons,
and a strong non-thermal behavior even at $E/W^*\sim 0.1$.
To evaluate $T_{\rm eff}$, fit to a Fermi-Dirac function with $T_{\rm eff}$ has
been performed on data satisfying $|f_{\rm loc}(\omega)-0.5|<0.25$.

Even though the calculations performed here are on homogeneous lattices,
the phase coexistence suggests that, under a uniform field, the system
can be spatially segregated into metal and insulator
regions which in turn have inhomogeneous temperature distribution with
complex thermodynamic states. The hot metallic regions will be oriented
in the direction of the field, forming experimentally observed
current-carrying filaments.

The Joule heating scenario has been previously invoked in the literature
for resistive switching in disordered films~\cite{altshuler}. Our
calculations of the coexistence of two distinct non-equilibrium
steady-state solutions in the framework of a relatively simple quantum
mechanical model could be applicable to NiO~\cite{noh} and
Cr$_x$V$_{2-x}$O$_3$~\cite{mcwhan} systems where metal-to-insulator
transitions occur with increasing temperature. Our
calculation ignores long-range anti-ferromagnetic correlations and does
not address switching from ordered insulating phases. Further extensions to
cluster-DMFT would allow a realistic treatment of the electronic
structure and could successfully address the case of VO$_2$.

The authors are grateful for helpful discussions with Satoshi Okamoto,
Sambandamurthy Ganapathy and Sujay Singh.  This work has been supported
by the National Science Foundation with the Grants No. DMR-0907150,
DMR-115181,  DMR-1308141, PHYS-1066293 and the hospitality of the Aspen
Center for Physics.


\begin{thebibliography}{*}
\bibitem{impurity} In a more realistic model with
impurity scattering which becomes more effective than the dissipation at
small fields, the critical field has different behavior $E_{\rm
lin}\propto \tau_{\rm imp}^{-1/2}\Gamma/U$.
\bibitem{lowT} At the metal-to-insulator RS, the
$T_{\rm eff}$ cools as far as the insulating state is allowed as a
meta-stable solution in the
equilibrium phase diagram, Fig.~\ref{fig4}(b). Therefore the $T_{\rm eff}$ at the $E$-field
immediately after the upper switching field $E_{\rm MIT}$ maps to the red line in Fig.~\ref{fig4}(b). 
After the RS to insulator, the current, while reduced by orders of magnitude,
self-consistently produces the Joule heat enough to support the insulating
solution. Also see Fig.~\ref{fig4}(c).
\end{thebibliography}
\end{document}


\myfonts

\centerline{\textbf{Supplementary Material: Metal-to-insulator phase
transition}}
\centerline{\textbf{ in field-driven electron lattice coupled to dissipative baths
}}
\bigskip
\centerline{Jiajun Li, Camille Aron, Gabriel Kotliar and Jong E. Han}

\bigskip

\begin{figure}[h]
\rotatebox{0}{\resizebox{!}{1.7in}{\includegraphics{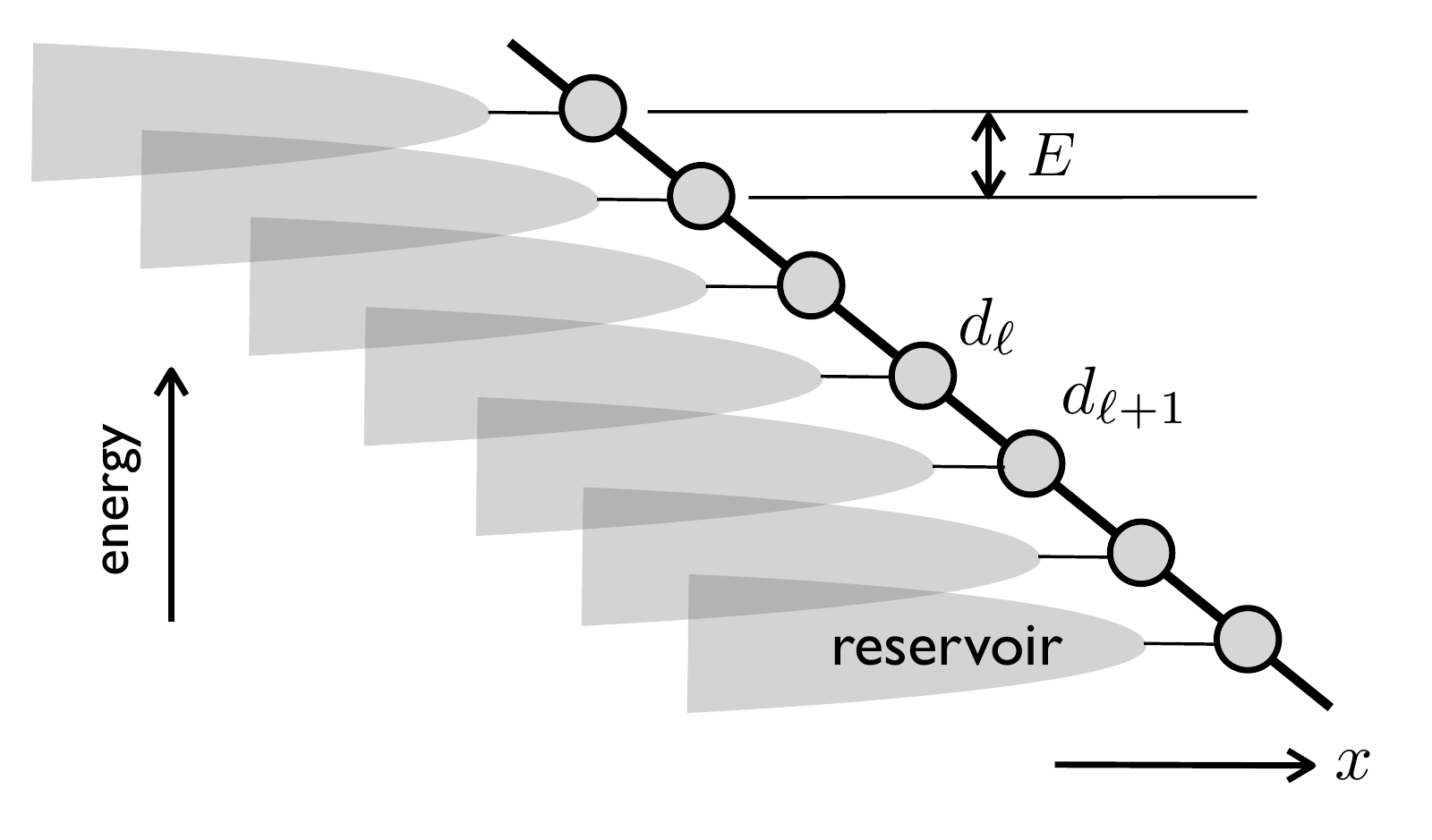}}}
\caption{One-dimensional tight-binding chain driven by a uniform
electric-field within the Coulomb gauge. Each orbital along the
transport chain is connected to a semi-infinite fermionic chain which
dissipates the excess energy accumulated by the Joule heating.
}
\label{figdia}
\end{figure}

\section{Formulation of one-dimensional chain under
electric-field}
We model a one-dimensional Hubbard model in the Coulomb gauge, as shown
in Fig.~\ref{figdia}, with the Hamiltonian
\begin{eqnarray}
H & = & -\gamma\sum_{\ell\sigma}(d^\dagger_{\ell+1,\sigma}d_{\ell\sigma}+h.c.)+U\sum_\ell
\left(n_{\ell\uparrow}-\frac12\right)\left(n_{\ell\downarrow}-\frac12\right) + \sum_{\ell\alpha\sigma}\epsilon_\alpha
c^\dagger_{\ell\alpha\sigma}c_{\ell\alpha\sigma}\nonumber \\
& & -\frac{g}{\sqrt{L}}\sum_{\ell\alpha\sigma}
(d^\dagger_{\ell\sigma}c_{\ell\alpha\sigma}+h.c.)
-\sum_{\ell\sigma}\ell E\left(
d^\dagger_{\ell\sigma}d_{\ell\sigma}+\sum_\alpha
c^\dagger_{\ell\alpha\sigma}c_{\ell\alpha\sigma}\right),
\end{eqnarray}
where all orbitals on the $\ell$-th TB site, and their chemical potential, are shifted by $\ell E$. Here we use the
unit $e=a=1$. The fermionic chain reservoirs drain the excess energy of the excited electrons on the main tight-binding (TB) lattice. 
This Hamiltonian in the Coulomb gauge~\cite{diss2} is equivalent to
the temporal gauge~\cite{turkowski} and the gauge-covariant
form~\cite{aron}. In a long-time limit, we assume we have already reached
a well-defined nonequilibrium steady state in the presence of reservoirs. With respect
to $d_\ell$, we have two sources for the electronic self-energy, one from the
fermion reservoirs
and the other from the Coulomb interaction, which we denote
as $\Sigma_\Gamma$ and $\Sigma_U$, respectively. Here we
make a dynamical mean-field theory (DMFT) assumption that the
self-energies are local and identical, except for the energy shift due to the voltage drop along the TB
lattice~\cite{satosh},
\begin{equation}
G^{r,<}_{\ell\ell}(\omega)=G^{r,<}_{\rm loc}(\omega+\ell E)
\mbox{ and }
\Sigma^{r,<}_{\ell\ell}(\omega)=\Sigma^{r,<}_{\rm loc}(\omega+\ell E),
\end{equation}
$\ell=-\infty,\infty$ denotes the lattice site~\cite{diss2}. Here the subscript `loc'
refers to the local quantity at the central site $\ell=0$. The Dyson's equation in
the steady state for the full retarded
Green's function can be expressed in the familiar form as
\begin{eqnarray}
{\bf
G}^r(\omega)^{-1} & = &
\left[
\begin{array}{ccccc}
\ddots & & & & \\
& \omega-E+\rmi\Gamma-\Sigma^r_{U,{\rm loc}}(\omega-E) & \gamma & 0 & \\
& \gamma & \omega+\rmi\Gamma-\Sigma^r_{U,{\rm loc}}(\omega) & \gamma &  \\
& 0 & \gamma & \omega+E+\rmi\Gamma-\Sigma^r_{U,{\rm loc}}(\omega+E) & \\
& & & & \ddots 
\end{array}
\right]\nonumber \\
& = & [\omega+\ell E+\rmi\Gamma-\Sigma^r_{U,{\rm loc}}(\omega+\ell
E)]\delta_{\ell\ell'}+\gamma\delta_{|\ell-\ell'|,1}.
\end{eqnarray}
The Weiss-field Green function ${\cal G}$ can be expressed similarly
except that the interacting self-energy is omitted at the central site,
\begin{equation}
[{\cal G}^r(\omega)^{-1}]_{\ell\ell'}=[{\bf
G}^r(\omega)^{-1}]_{\ell\ell'}+\Sigma^r_{U,{\rm loc}}(\omega)\delta_{\ell
0}\delta_{\ell' 0}\equiv[{\bf
G}^r(\omega)^{-1}+\bm{\Sigma}^r_{U,{\rm loc}}]_{\ell\ell'},
\end{equation}
with $\bm{\Sigma}^r_{U,{\rm loc}}={\rm
diag}[\cdots,0,0,\Sigma^r_{U,{\rm loc}}(\omega),0,0,\cdots]$.

The inversion of the above infinite matrix can be achieved efficiently by a
recursive method. We divide the lattice into three parts with the
central site $\ell=0$, the left ($\ell=-1,-2,\cdots,-\infty$)
and right ($\ell=1,2,\cdots,\infty$) semi-infinite chains. We denote the
retarded GF matrix ${\cal F}^r_+$ on the RHS semi-infinite chain as
\begin{equation}
[{{\cal F}^r_+(\omega)}^{-1}]_{\ell\ell'}=
[\omega+\ell E+\rmi\Gamma-\Sigma^r_{U,{\rm loc}}(\omega+\ell
E)]\delta_{\ell\ell'}+\gamma\delta_{|\ell-\ell'|,1},
\end{equation}
with $\ell,\ell'=1,2,\cdots,\infty$. The local GF at the end of the
chain ($\ell=1$) [$F^r_+(\omega+E)\equiv{\cal
F}^r_+(\omega)_{11}$] can be expressed as a continued fraction
\begin{eqnarray}
F^r_+(\omega+E) & = & \frac{1}{\omega+E+\rmi\Gamma-\Sigma^r_{U,{\rm
loc}}(\omega+E)-\frac{\gamma^2}{\omega+2E+\rmi\Gamma-\Sigma^r_{U,{\rm
loc}}(\omega+2E)-\frac{\gamma^2}{\cdots}}}\nonumber \\
& = & [\omega+E+\rmi\Gamma-\Sigma^r_{U,{\rm loc}}(\omega+E)
-\gamma^2 F^r_+(\omega+2E)]^{-1},\\
\mbox{ or }
F^r_+(\omega+E)^{-1} & = & \omega+E+\rmi\Gamma-\Sigma^r_{U,{\rm
loc}}(\omega+E) -\gamma^2 F^r_+(\omega+2E),
\label{eq:recur1}
\end{eqnarray}
from the self-similarity of the semi-infinite chain. The recursive
relation Eq.~(\ref{eq:recur1}) is solved numerically with iteration
number $M$ over 500. Practically, we start from an
initial GF $F^r_+(\omega+ME)
=[\omega+ME+\rmi\Gamma-\Sigma^r_{U,{\rm loc}}(\omega+ME)]^{-1}$ and by
Eq.~(\ref{eq:recur1}) we generate $F^r_+(\omega+(M-1)E)$. We repeat the
process Eq.~(\ref{eq:recur1})
until we reach $F^r_+(\omega+E)$. The LHS GF, $F^r_-(\omega-E)$,
can be similarly obtained through
\begin{equation}
F^r_-(\omega-E)^{-1} = \omega-E+\rmi\Gamma-\Sigma^r_{U,{\rm
loc}}(\omega-E) -\gamma^2 F^r_-(\omega-2E).
\label{eq:recur2}
\end{equation}
Once we obtain fully convergent GFs $F^r_{\pm}(\omega\pm E)$, the full
local GF for the infinite chain can be constructed as
\begin{equation}
G^r_{\rm loc}(\omega)^{-1}=\omega+\rmi\Gamma-\Sigma^r_{U,{\rm loc}}(\omega)
-\gamma^2[F^r_+(\omega+E)+F^r_-(\omega-E)].
\label{eq:gr}
\end{equation}
The Weiss-field GF ${\cal G}^r(\omega)$, omits the interacting self-energy only on the central
site ($\ell=0$) and we have
\begin{equation}
{\cal G}^r(\omega)^{-1}=\omega+\rmi\Gamma-\gamma^2[F^r_+(\omega+E)+F^r_-(\omega-E)]
=G^r_{\rm loc}(\omega)^{-1}+\Sigma^r_{U,{\rm loc}}(\omega).
\label{eq:wfr}
\end{equation}

Now, we turn to the Dyson's equation for lesser GFs.
When the lattice of $d_\ell$ is connected to the reservoirs and
with finite interaction, its steady-state dynamics is governed by the 
transport equation
\begin{equation}
G^<_{\ell\ell'}(\omega)=
\sum_{p}G^r_{\ell p}(\omega)\Sigma^<_{p,{\rm tot}}(\omega)G^a_{p\ell'}(\omega),
\end{equation}
with $p=-\infty,\cdots,\infty$ running over all TB sites and
$\Sigma^<_{\rm tot}$ being the sum of contributions from the fermion
baths and the Hubbard interaction. For the central site $\ell=\ell'=0$, we use a 
similar trick as above
to group $p$ into the central site and left and right chains,
\begin{equation}
G^<_{\rm loc}(\omega)=G^r_{\rm loc}(\omega)\Sigma^<_{\rm tot,loc
}(\omega)G^a_{\rm loc}(\omega)+
\sum_{p<0}G^r_{0p}(\omega)\Sigma^<_{{\rm
tot},p}(\omega)G^a_{p0}(\omega)
+\sum_{p>0}G^r_{0p}(\omega)\Sigma^<_{{\rm
tot},p}(\omega)G^a_{p0}(\omega).
\end{equation}
For the RHS summation ($p>0$), one can write the Dyson's equation
$G^r_{0p}(\omega)=G^r_{\rm loc}(\omega)(-\gamma){\cal
F}^r_{+,1p}(\omega)$ and similarly for the advanced FGs, and therefore we have the
third term as
\begin{equation}
\gamma^2|G^r_{\rm loc}(\omega)|^2\sum_{p>0}{\cal F}^r_{+,1p}(\omega)\Sigma^<_{{\rm
tot},p}(\omega){\cal F}^a_{+,p1}(\omega).
\end{equation}
The summed expression is nothing but the local lesser GF
$F^<_+(\omega+E)={\cal F}^<_{+,11}(\omega)$
within the LHS semi-infinite chain, and we obtain
\begin{equation}
G^<_{\rm loc}(\omega)=|G^r_{\rm loc}(\omega)|^2\left\{\Sigma^<_{\rm tot,loc
}(\omega)+\gamma^2[F^<_+(\omega+E)+F^<_-(\omega-E)]\right\}.
\label{eq:glss}
\end{equation}
$F^<_\pm(\omega\pm E)$ can be obtained from $\Sigma^<_{\rm loc}(\omega)$
following similar steps.
\begin{eqnarray}
F^<_+(\omega+E)&=&\sum_{p=1}^\infty {\cal F}^r_{+,1p}(\omega)
\Sigma^<_{{\rm tot},p}(\omega){\cal F}^a_{+,p1}(\omega)\nonumber \\
&=&|F^r_+(\omega+E)|^2\Sigma^<_{\rm tot,1}(\omega)
+\gamma^2|F^r_+(\omega+E)|^2\sum_{p=2}^\infty
\tilde{\cal F}^r_{+,2p}(\omega)
\Sigma^<_{{\rm tot},p}(\omega)\tilde{\cal F}^a_{+,p2}(\omega),\nonumber
\end{eqnarray}
where the tilde denotes that the GFs are on the semi-infinite chains
of $p=2,3,\cdots,\infty$. Using the self-similarity of the chains
$\ell=1,\cdots,\infty$ and $\ell=2,\cdots,\infty$, we have
\begin{equation}
F^<_\pm(\omega\pm E)=|F^r_\pm(\omega\pm E)|^2\left[\Sigma^<_{\rm tot,loc}
(\omega\pm E)+\gamma^2F^<_\pm(\omega\pm 2E)\right].
\label{eq:flss}
\end{equation}
The lesser Weiss-field GF can be written as
\begin{equation}
{\cal G}^<(\omega)=|{\cal G}^r(\omega)|^2\left\{\Sigma^<_{\Gamma,{\rm
loc}}
(\omega)+\gamma^2[F^<_+(\omega+E)+F^<_-(\omega-E)]\right\},
\end{equation}
with the damping part of the self-energy $\Sigma_\Gamma$.
Using Eq.~(\ref{eq:glss}),
\begin{equation}
{\cal G}^<(\omega)  =  |{\cal G}^r(\omega)|^2\left(
\frac{G^<_{\rm loc}(\omega)}{|
G^r_{\rm loc}(\omega)|^2}-\Sigma^<_{U,{\rm loc}}(\omega)\right).
\label{eq:wflss}
\end{equation}
In the main paper, the subscript `loc' has been omitted for brevity.
Electric current per spin is calculated as
\begin{eqnarray}
J &=& \frac{i}{2} \gamma\langle d^\dagger_{1\sigma}d_{0\sigma}
-d^\dagger_{0\sigma}d_{1\sigma}
+d^\dagger_{0\sigma}d_{-1\sigma}
-d^\dagger_{-1\sigma}d_{0\sigma}\rangle
 =  \gamma{\rm Re}[G^<_{01}(t=0)-G^<_{0-1}(t=0)]\nonumber \\
& = & \gamma{\rm Re}\int \frac{d\omega}{2\pi}
[G^<_{01}(\omega)-G^<_{0-1}(\omega)] \\
& = & -\gamma^2{\rm Re}\int \frac{d\omega}{2\pi}
\left\{G^<(\omega)\left[F^a_+(\omega+\Omega)-F^a_-(\omega-\Omega)\right]
+G^r(\omega)\left[F^<_+(\omega+\Omega)-F^<_-(\omega-\Omega)\right]\right\}.
\nonumber
\end{eqnarray}

To summarize, given the local self-energies $\Sigma^{r,<}_{\rm
loc}(\omega)$, GFs for the semi-infinite chains $F^{r,<}_{\pm}$ 
are calculated via
Eqs.~(\ref{eq:recur1},\ref{eq:recur2},\ref{eq:flss}). The
retarded GFs are obtained via Eqs.~(\ref{eq:gr},\ref{eq:wfr}), and
finally the lesser GFs follow via Eqs.~(\ref{eq:glss},\ref{eq:wflss}).
This procedure, formulated on real-space, corresponds to the
$k$-summation of the impurity GF in equailibrium DMFT formalism.

\medskip
\noindent\textbf{Multi-dimensional lattice under
electric-field:}
For higher dimensional cubic lattice with the field along an
axial direction (${\bf E}=E\hat{\bf x}$), the lattice has translational
invariance perpendicular to the field, and the problem is 
block-diagonalized with the transverse wave-vector ${\bf k}_\perp$.
We solve the Dyson's equation as above with the ${\bf
k}_\perp$-space (the self-energy $\Sigma^{r,<}_{\rm loc}(\omega)$ does
not have ${\bf k}_\perp$ dependence), and then sum over ${\bf k}_\perp$ to get the local GF. For
hypercubic TB lattice the dispersion is $\epsilon_{\bf k}=-2\gamma\cos(k_x)
+\epsilon({\bf k}_\perp)$. Then
adding $\epsilon({\bf k}_\perp)$ to the on-site energy
of the 1-$d$ tight-binding chain and
carrying out the 1-$d$ Dyson's equation in the previous section, we
obtain the GF $G^{r,<}_{{\bf k}_\perp}(\omega)$. By summing over
${\bf k}_\perp$ in the $d-1$ dimensional Brillouin zone,
we get the full local GFs
\begin{equation}
G^{r,<}_{\rm loc}(\omega)=\int_{\rm BZ}\frac{d^{d-1}{\bf
k}_\perp}{(2\pi)^{d-1}}G^{r,<}_{{\bf k}_\perp}(\omega)
=\int d\epsilon_\perp D_{d-1}(\epsilon_\perp)G^{r,<}(\epsilon_\perp,\omega),
\end{equation}
with the $d-1$ dimensional DoS $D_{d-1}(\epsilon_\perp)$.
The Weiss-field GFs are obtained via
Eqs.~(\ref{eq:wfr},\ref{eq:wflss}).

\section{Crossover of effective temperature near $U_{\rm cross}\approx
1.32W$} The effective temperature $T_{\rm eff}$ in Fig. 4(c) shows
different behavior around $U_{\rm cross}\approx 1.32W$, where $U<U_{\rm
cross}$ the increase of $T_{\rm eff}$ with the E-field is rapid in a
similar fashion as that well away from the coexisitence region with
smaller $U$, whereas for $U>U_{\rm cross}$ the increase of $T_{\rm eff}$
is much slower after the linear response limit. Therefore it leads to a
large electric-field to reach the metal-insulator transition near the
crossover, resulting in a maximum of the $E_{\rm MIT}(U)$ curve in
Fig.~4(a) and (c). Here, we give a sketch of the different scaling
behaviors in the two regimes.

From the balance of the Joule heating and the dissipation of energy into
the fermion baths, we arrive at the rigorous relation Eq.~(39) of Han
and Li~\cite{diss2},
\begin{equation}
JE=2\Gamma\int\omega A(\omega)[f_{\rm loc}(\omega)-f_{\rm
b}(\omega)]d\omega,
\label{balance}
\end{equation}
where $A(\omega)$ is the on-site spectral function of the tight-binding
lattice, $f_{\rm b}(\omega)$ the Fermi-Dirac function of the bath. We
approximate the local distribution function $f_{\rm loc}(\omega)$ as a
Fermi-Dirac function with the effective temperature $T_{\rm eff}$.

(i) In the regime of $U<U_{\rm cross}$ with $W^*>T_{\rm eff}\gg T_{\rm b}$, we can apply
the Sommerfeld expansion and obtain
\begin{equation}
JE\approx \frac{\pi^2}{3}\Gamma A(0)(T_{\rm eff}^2-T_{\rm b}^2),
\label{phenom}
\end{equation}
which agrees with the phenomenological energy balance equation of
Altshuler et al~\cite{altshuler} with the dissipation given by the fermion baths. Away from the
linear response limit, $\tau_U^{-1}\gg\Gamma$ and $J\propto
\gamma E/\tau_U^{-1}$ and we have
\begin{equation}
E^2\propto \Gamma \tau_U^{-1}T_{\rm eff}^2/W^2.
\end{equation}
From the perturbative self-energy for the scattering rate, as used in
the main text,
\begin{equation}
\tau_{U}^{-1} = -{\rm Im}\Sigma^r_{\rm eq}(\omega=0,T_{\rm eff})
\approx \frac{\pi^3}{2} \, A_0(0)^3 \, U^2 \, T_{\rm eff}^2,
\label{eq:scatt}
\end{equation}
we arrive to the scaling relation,
\begin{equation}
E^2\propto \Gamma U^2T_{\rm eff}^4/W^5,\mbox{ or }
T_{\rm eff}/W\propto (E/U)^{1/2},
\label{relations}
\end{equation}
with $T_{\rm eff}$ increasing as $\sqrt{E}$ beyond the linear response
regime. This scaling agrees well with the numerical calculations.

(ii) In the regime of $U>U_{\rm cross}$ with $T_{\rm eff}\lesssim W^*$, the scaling
relation is quite different. Although somewhat exaggerated,
we assume $T_{\rm eff}\gg W^*$ in the following argument
for the sake of simplicity. In such limit, the
Sommerfeld expansion is not applicable to both the Eqs.~(\ref{phenom})
and (\ref{eq:scatt}), and the half-QP-bandwidth $W^*/2$ replaces the
role of temperature $\pi T_{\rm eff}$,
leading to the approximate relations
\begin{equation}
JE\propto (\Gamma/W){W^*}^2\mbox{ and }
\tau_{U}^{-1} \propto U^2 {W^*}^2/W^3,
\end{equation}
which demonstrates that $T_{\rm eff}$-dependence effectively drops out.
Therefore, as shown in the numerical calculations in Fig. 4(c),
$T_{\rm eff}$ has much reduced dependency on E-field for $U>U_{\rm
cross}$, away from the linear response regime. This slow increase of $T_{\rm
eff}$ leads to
the enhenced upper switching field $E_{\rm MIT}(U)$ to reach the resistive switching, 
and the maximum behavior of $E_{\rm MIT}(U)$ results near the crossover value
$U\approx U_{\rm cross}$.

\begin{figure}[h]
\rotatebox{0}{\resizebox{!}{1.7in}{\includegraphics{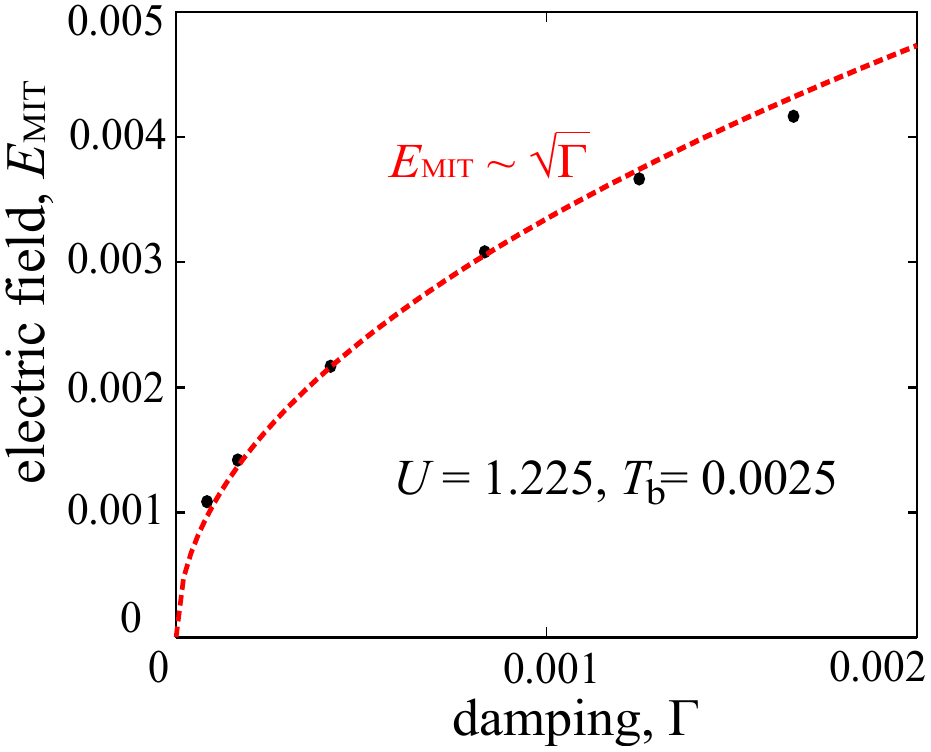}}}
\caption{Scaling relation of the critical field for the resistive
switching vs. the damping parameter $\Gamma$, as the electric-field is
increased. The relation relation
$E_{\rm MIT}(\Gamma)\propto\sqrt{\Gamma}$ is consistent with the thermal
scenario.
}
\label{figscale}
\end{figure}

The above scaling relations can be used to derive $E_{\rm
MIT}(U)$'s dependence on $\Gamma$. At a given $U$, the nonequilibrium
MIT occurs when the effective temperature matches
the equilibrium transition temperature $T_{\rm eff}=T_{\rm eq,MIT}(U)$.
Then Eq.~(\ref{relations}), for $U<U_{\rm cross}$ leads to
\begin{equation}
E_{\rm MIT}(U)^2\propto\Gamma U^2T_{\rm eq,MIT}^4/W^5,
\mbox{ and }
E_{\rm MIT}(U)\propto\sqrt\Gamma.
\end{equation}
The numerical calculations for $U=1.225$ shown in Fig.~\ref{figscale} confirms the
relation.